\documentclass{emulateapj}
\journalinfo{\@To be submitted to the Astrophysical Journal}

\shorttitle{The Dwarfs Beyond: The Stellar-to-Halo Mass Relation}
\shortauthors{Miller et al.}

\begin{document}

\title{The Dwarfs Beyond: The Stellar-to-Halo Mass Relation \\ for a New Sample of Intermediate Redshift Low Mass Galaxies}

\author{
Sarah H. Miller\altaffilmark{1,2,3},
Richard S. Ellis\altaffilmark{1},
Andrew B. Newman\altaffilmark{1,4}, \&
Andrew Benson\altaffilmark{4}
}

\email{smiller@astro.caltech.edu}

\altaffiltext{1}{California Institute of Technology, 1200 E. California Blvd, Pasadena, CA 91125}
\altaffiltext{2}{University of California - Riverside, 900 University Ave, Riverside, CA 92521}
\altaffiltext{3}{University of California - Irvine, Irvine, CA 92697}
\altaffiltext{4}{Carnegie Observatories, 813 Santa Barbara St, Pasadena, CA 91101}

\begin{abstract}

A number of recent challenges to the standard $\Lambda$-CDM paradigm relate to discrepancies that arise in comparing the abundance and kinematics of local dwarf galaxies with the predictions of numerical simulations. Such arguments rely heavily on the assumption that the Local Volume's dwarf and satellite galaxies form a representative distribution in terms of their stellar-to-halo mass ratios. To address this question, we present new, deep spectroscopy using DEIMOS on Keck for 82 low mass ($10^7$-$10^9$ M${}_{\odot}$) star-forming galaxies at intermediate redshift ($0.2 < z < 1$). For 50\% of these we are able to determine resolved rotation curves using nebular emission lines and thereby construct the stellar mass Tully-Fisher relation to masses as low as $10^7$ M$_{\odot}$. Using scaling relations determined from weak lensing data, we convert this to a stellar-to-halo mass (SHM) relation for comparison with abundance matching predictions. We find a discrepancy between the propagated predictions from simulations compared to our observations, and suggest possible reasons for this as well as future tests that will be more effective.

\end{abstract}

\section{Introduction}

The $\Lambda$-dominated Cold Dark Matter model ($\Lambda$CDM) has been remarkably successful in the interpretation of the large scale structure of the Universe and its evolution as probed by observations of the cosmic microwave background (CMB), the local galaxy distribution and independent probes of the dark matter power spectrum such as the Lyman-$\alpha$ forest and weak gravitational lensing. However, significant challenges remain when the theory is confronted with observations on galaxy scales \citep[see review by][]{weinbe2013}.

A long-standing question is the apparent mismatch of the abundance of visible satellites in the Milky Way halo compared to that predicted from the steep dark matter power spectrum \citep{klypin1999,moore1999}. Possible explanations include the likelihood that halos are dark due to the early photoionizing background \citep{bulloc2000}, energetic feedback which suppressed growth \citep{font2011}, or an observational bias whereby many satellites have yet to be discovered, either because they are hidden by the Galactic plane or are too faint for existing surveys \citep{toller2008}. Recently, a more fundamental challenge relates to a surprising discrepancy between the observed and expected maximum circular velocities ($V_{max}$) for the most massive Local Group satellites. The best-studied dwarf spheroidals near the Milky Way have $12< V_{max} <25 $ km s${}^{-1}$ whereas the Aquarius \citep{spring2008} and the Via Lactea II simulations \citep{dieman2007, dieman2008} predict at least ten subhaloes should be visible with $V_{max} > 25$ km s${}^{-1}$ \citep{boylan2011}. These simulations predict DM haloes five times more massive than what could be inferred given the observed densities of the satellite dwarf spheroidals. No incontrovertible explanation currently satisfies the ``Too Big to Fail" \citep{boylan2011} problem; some consider it a result of poorly-understood baryonic feedback effects \citep[e.g.,][]{pontze2012}, whereas others postulate a fundamental departure from the current dark matter model \citep[e.g.,][]{rocha2013}.

Both of the interrelated problems above rely inescapably on the question of the whether the Milky Way and its satellites are representative of those typical of the larger cosmic volumes. Furthermore, predicting satellite properties of the Milky Way is complicated by the difficulty in inferring its halo mass. For dwarf galaxies and satellites, the halo mass range $10^7 M_{\odot}< M < 10^9 M_{\odot}$, where resolved kinematic data are sparse, is particularly interesting. A current inventory reveals rotation curves for both Magellanic Clouds and 9 of the brightest satellites of the Milky Way. The HI Nearby Galaxy Survey (THINGS) contributes a further 4 galaxies \citep{oh2011} to the 8 gas-rich systems from \citet{stark2009}. 

A more general way to view this problem is the relationship between the stellar mass and 
halo mass (the so-called SHM relation). This relation is well-constrained down to stellar masses of $10^9$ M${}_{\odot}$ by various methods, including weak lensing \citep[e.g.,][]{mandel2006,leauth2012}, abundance matching \citep[e.g.,][]{moster2010,behroo2010} and stellar kinematics \citep[e.g.,][]{conroy2007,more2011}. The most effective probe for lower mass systems is resolved dynamics for rotationally-supported systems, i.e. the stellar mass Tully-Fisher relation \citep[e.g.,][]{geha2006,pizagn2007}. Various methods can be used to relate observed velocities to the halo virial circular velocity, which we discuss further as motivation for this work in the following section.

This paper presents results from a new observational program whose goal is to extend the stellar mass Tully-Fisher relation into the relevant mass range $10^7$ M${}_{\odot} <$ M${}_{\ast} < 10^9$ M${}_{\odot}$ at intermediate redshifts, and constrain the stellar-to-halo mass (SHM) relation well beyond the confines of the Local Group. As we demonstrate here, such an observational program to investigate the dwarf density discrepancy described here is now feasible. 

A plan of the paper is as follows: in \S 2 we describe our motivation in evaluating the SHM relation in the dwarf regime. In \S 3, we describe the selection criteria for our sample, the DEIMOS spectroscopic data and the resolved photometry from the Hubble Space Telescope (HST) from which our stellar mass estimates are derived. Specifically \S 3.4 discusses the techniques we use to fit kinematic models to our low mass galaxies. In \S 4, we present our stellar mass Tully-Fisher relation and introduce our method for converting this relation to the required stellar-to-halo mass (SHM) relation. In \S 5 we discuss our SHM results in the context of predictions from $\Lambda$CDM cosmological simulations and abundance matching methods. Throughout the paper we adopt a $\Omega_{\Lambda}$ = 0.7, $\Omega_{m}$ = 0.3, $H_0$ = 70 km sec$^{-1}$ Mpc$^{-1}$ cosmology. All magnitudes refer to the AB system.

\section{Motivation}

The methods used for relating velocities measured across the optical extent of galaxies to the halo virial circular velocity, or $V_{200}$ \citep[e.g.,][]{navarr1997,bulloc2001,reyes2012}, have not been calibrated directly in the dwarf regime. Yet if the relation of \citet{reyes2012}\footnote{In \citet{reyes2012}, the optical-to-virial velocity ratios are derived from both directly measured velocities from rotation curves along with $V_{200}$ values from halo masses derived with galaxy-galaxy weak lensing. A consistent and calibrated selection was used between samples rotation curve and weak lensing samples, and the constructed Tully-Fisher relations are consistent between their work at $z\sim0$ and our previous work \citep{miller2011} with the same mass range at higher redshift.}  to convert optical velocities to halo velocities is extrapolated to low mass, there is a serious discrepancy between the resulting halo circular velocities and those calculated based on abundance matching techniques for dwarf galaxies $<10^9$ M${}_{\odot}$ (Fig. 1). 

A key question is whether this offset is a generic result for low mass dwarf galaxies beyond the Local Group. Furthermore, how do the Local Group galaxies relate to the full scatter in the stellar-to-halo mass relation of a larger, more representative sample? Although the conversion from dynamical mass to halo mass will require careful evaluation, given such scaling relations are currently only determined for higher mass systems than the focus of this work, data can hopefully provide the basis for future comparisons with numerical simulations. Nevertheless, while extrapolation of \citet{reyes2012} is uncertain, recent work in the Next Generation Virgo Survey suggests this extrapolation to be roughly accurate (Grossauer et al. in prep.).

\begin{figure}
\includegraphics[width=3.45in]{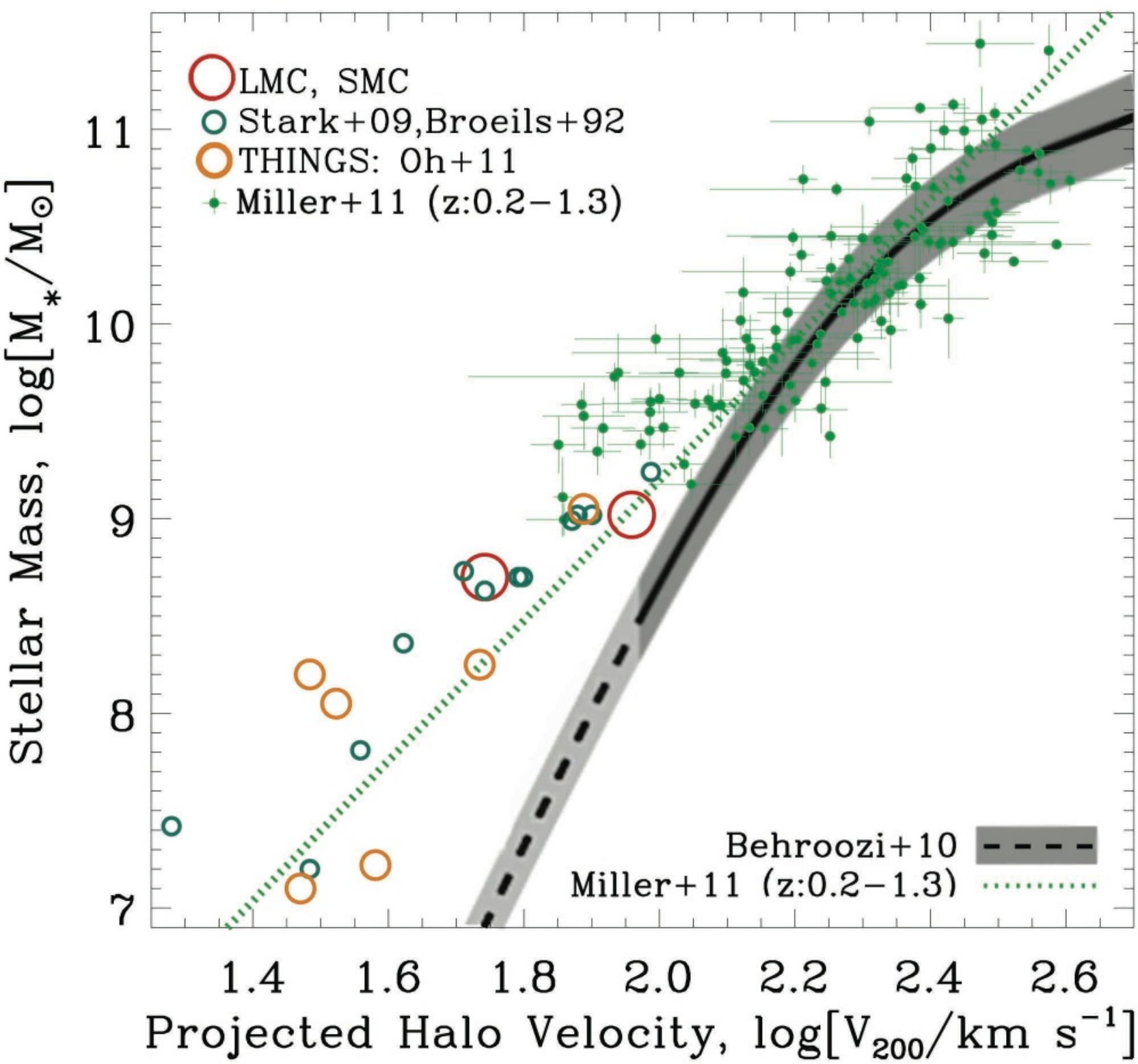}
\caption{Motivating this work, there appears to be a divergence between the abundance matching result and the stellar mass Tully-Fisher relation, which has been converted to halo velocity $V_{200}$ using the extrapolated \citet{reyes2012} relation between observed velocities as a function of stellar mass and $V_{200}$. Also plotted are the converted velocities of local group galaxies, which further suggest a discrepancy between observed velocities and abundance matching, however the scatter in the stellar-to-halo mass relation is unclear with so few data points.}
\label{fig:motivation}
\end{figure}

\begin{figure}
\includegraphics[width=3.5in]{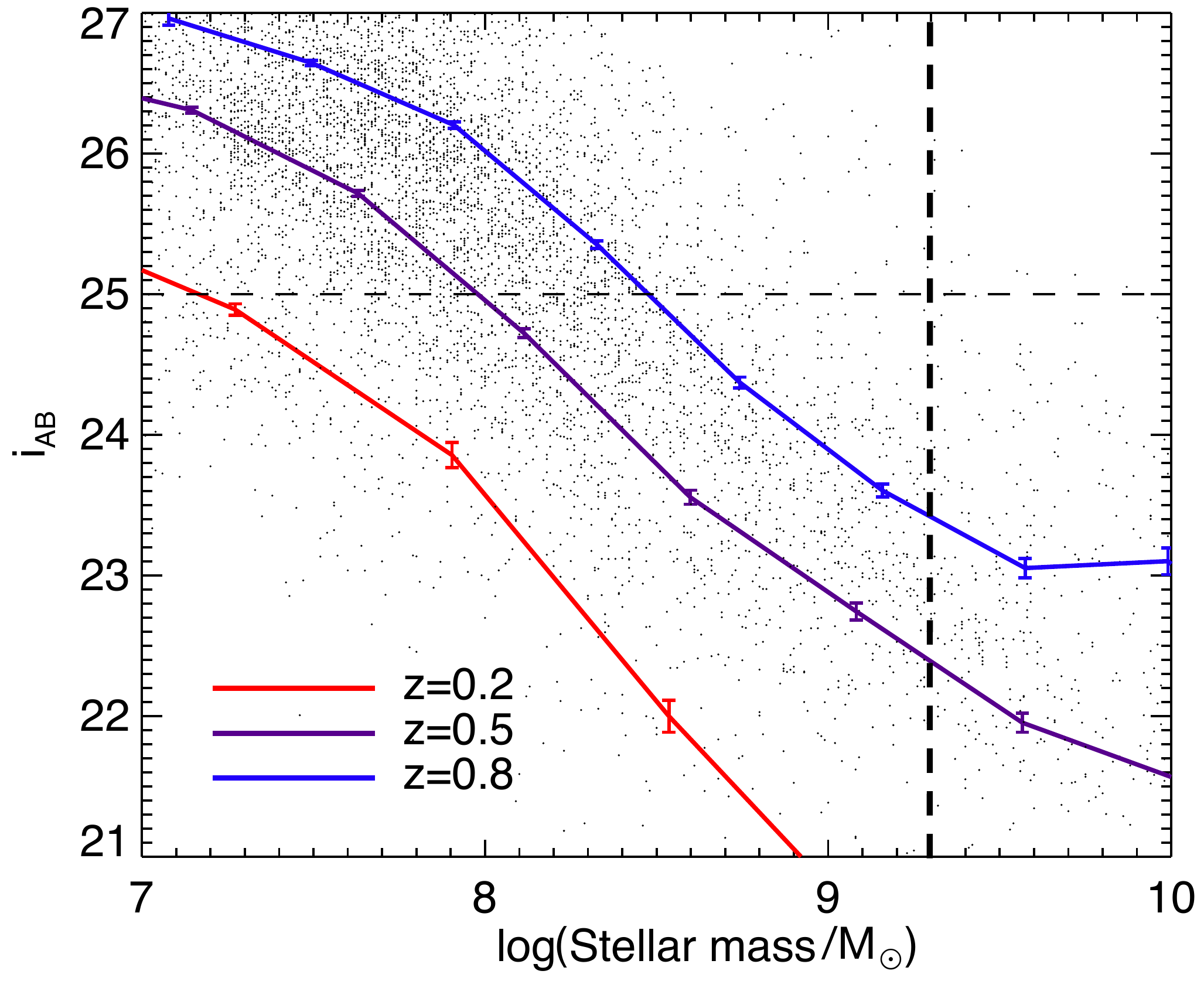}
\caption{Median magnitude curves within the \citet{newman2012} catalog as a function of apparent $i_{AB}$ magnitude for various redshifts of interest. For the redshift range $0.2< z <0.8$ galaxies with stellar masses as low as $10^{7.2}$ to $10^{8.5}$ M${}_{\odot}$, respectively, can be identified, thereby sampling the important range of the stellar-to-halo mass relation. Dashed lines denote the magnitude and stellar mass limits used in preparing the DEIMOS target list.}
\label{fig:maglimit}
\end{figure}

\section{Data}

We present a new spectroscopic dataset which exploits the significant multiplex gain of DEIMOS to address the kinematic properties of low mass star-forming dwarf galaxies at intermediate redshift ($0.2< z <1.0$). 
Two advances make it practical to target dwarf galaxies at intermediate redshifts: First is our demonstrated ability to recover rotational velocities on small angular scales from our earlier work on high redshift disk galaxies \citep{miller2011,miller2012a}, the relevant techniques of which we describe further in \S 3.4. Second is the availability of remarkably deep Hubble Space Telescope (HST) Wide-Field Camera 3 (WFC3) infra-red imaging from the Cosmic Assembly Near-IR Deep Extragalactic Legacy Survey \citep[CANDELS;][]{grogin2011,koekem2011}, enabling us to conduct resolved photometric models of dwarf targets and construct catalogs \citep[i.e.,][]{newman2012} of photometric redshifts and stellar mass estimates along with deep optical Subaru SXDS data for much lower mass systems at intermediate redshifts than previously possible. Although a 10 meter class telescope and long exposures are necessary to study these low mass sources efficiently, as the target density within a DEIMOS field is substantial, the multiplex gain offers a huge advantage over modest samples of dwarf galaxies studied individually in nearby galaxies. Targets of similar mass available from SDSS have a much lower surface density and cannot be efficiently surveyed with the current multiplex gains on 4- or 10-meter aperture telescopes.

\begin{figure}
\includegraphics[width=3.6in]{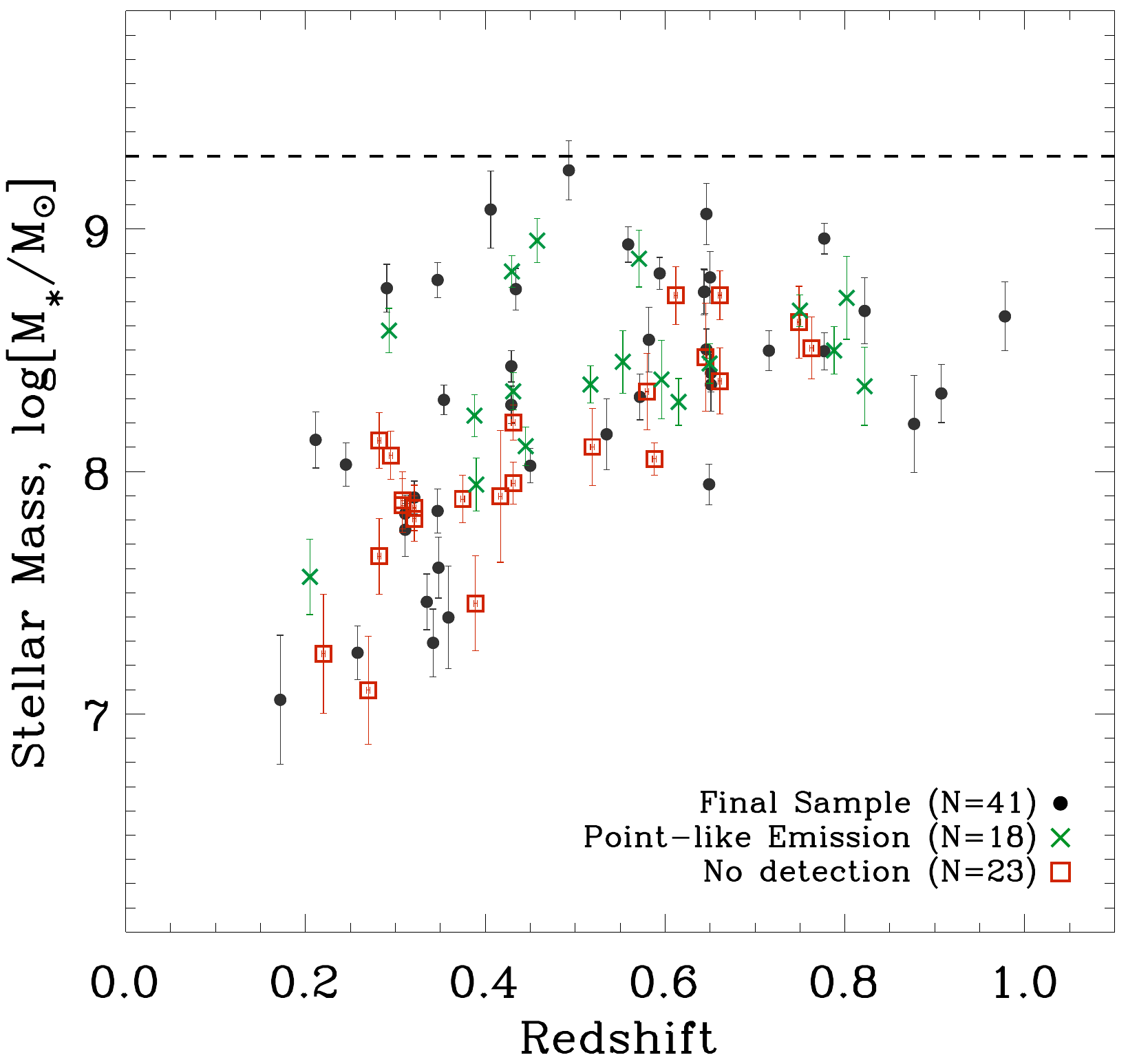}
\caption{The stellar mass and redshift distribution for the 82 dwarf galaxies studied spectroscopically. 23 galaxies which revealed no significant line emission are plotted as red boxes (in the colored version) at their photometric redshifts. 18 galaxies which revealed unresolved line emission are denoted with green crosses (in the colored version) at their spectroscopic redshift. The other 41 galaxies show resolved line emission are are denoted with black circular points. The dashed line marks the stellar mass limit for the overall sample.}
\label{fig:sample}
\end{figure}

\subsection{Sample Selection and DEIMOS Observations}

In this work we examine a large number of low mass galaxies at intermediate redshift drawn from the UKIRT Ultradeep Survey field (UDS) using the photometric catalog derived by  \citet{newman2012}.
We adopted a magnitude limit of $i_{AB}=25$ which corresponds approximately to a median stellar mass limit of $10^{7.2}, 10^{8.0}$, and $10^{8.5}$ M${}_{\odot}$ for $z \simeq 0.2$, 0.5, and 0.8, respectively (Fig. \ref{fig:maglimit}). Two further criteria were used to select targets, a photometric redshift range $0.2<z<0.8$ and a stellar mass range $\log{(}$M${}_{\ast}$/M${}_{\odot})<$ 9.3. Photometric redshifts were determined using the EAzY code \citep{bramme2008} and stellar masses were estimated using the FAST code \citep{kriek2009}--- for more details see \S \ref{sec:stmass}. In constructing the multislit DEIMOS mask, slitlets were aligned to the major axes for significantly inclined targets \citep[based on photometry using SExtractor:][]{bertin1996}. The mask position angle was selected so as to optimally fill that portion of the UDS field with available WFC3 coverage.

We used the 1200 l mm${}^{-1}$ grating in DEIMOS centered at 7500 \AA~so that [OIII] and H$\beta$ line emission would be sampled for the full targeted redshift range, H$\alpha$ emission for $z\simeq0.2-0.4$ and [OII] for $z<0.6$. With $\sim$30 km s${}^{-1}$ kinematic resolution, our earlier work demonstrated the ability to recover rotation curves with characteristic velocities of 50 km s${}^{-1}$ \citep{miller2011,miller2012a}.
Our modeling code was well-tested using simulated data to rotational velocities as low as  30 km s${}^{-1}$. This corresponds to the characteristic maximal velocity for dwarf galaxies with stellar masses of $<10^6$ M${}_{\odot}$ in the abundance matching method, but to a stellar mass of $10^{7.3}$ M${}_{\odot}$ in the extrapolated stellar mass TF relation of \citet{miller2011}. Although our typical targets are only 1.5-2$\arcsec$ across, this corresponds to the extent of more massive systems successfully targeted at $z>$1 using our code \citep{miller2012a,miller2013}.

We successfully secured 10 hours of science-quality exposure time with the DEIMOS instrument on a single mask in the UDS field on December 11-12, 2012. A total of 82 dwarfs were targeted, but not all objects observed show ordered rotation or line emission. However we do not seek to construct a complete sample via this approach; of course such completeness is not even possible in local data. In summary we find that 28\% (N=23) of the targeted sample reveal no significant line emission. These sources occupy the lower surface brightness portion of the photometric distribution suggesting their line emission may simply be too faint to be detectable.
We find 22\% (N=18) have unresolved line emission precluding any attempt to construct rotation curves. The remaining 50\% (N=41) have resolved emission from which we can attempt to derive rotation curves (see Fig. \ref{fig:sample}). This mix of target properties is actually similar to that recovered in our previous higher mass samples at intermediate redshifts.

\begin{figure*}
\center
\includegraphics[width=6.5in]{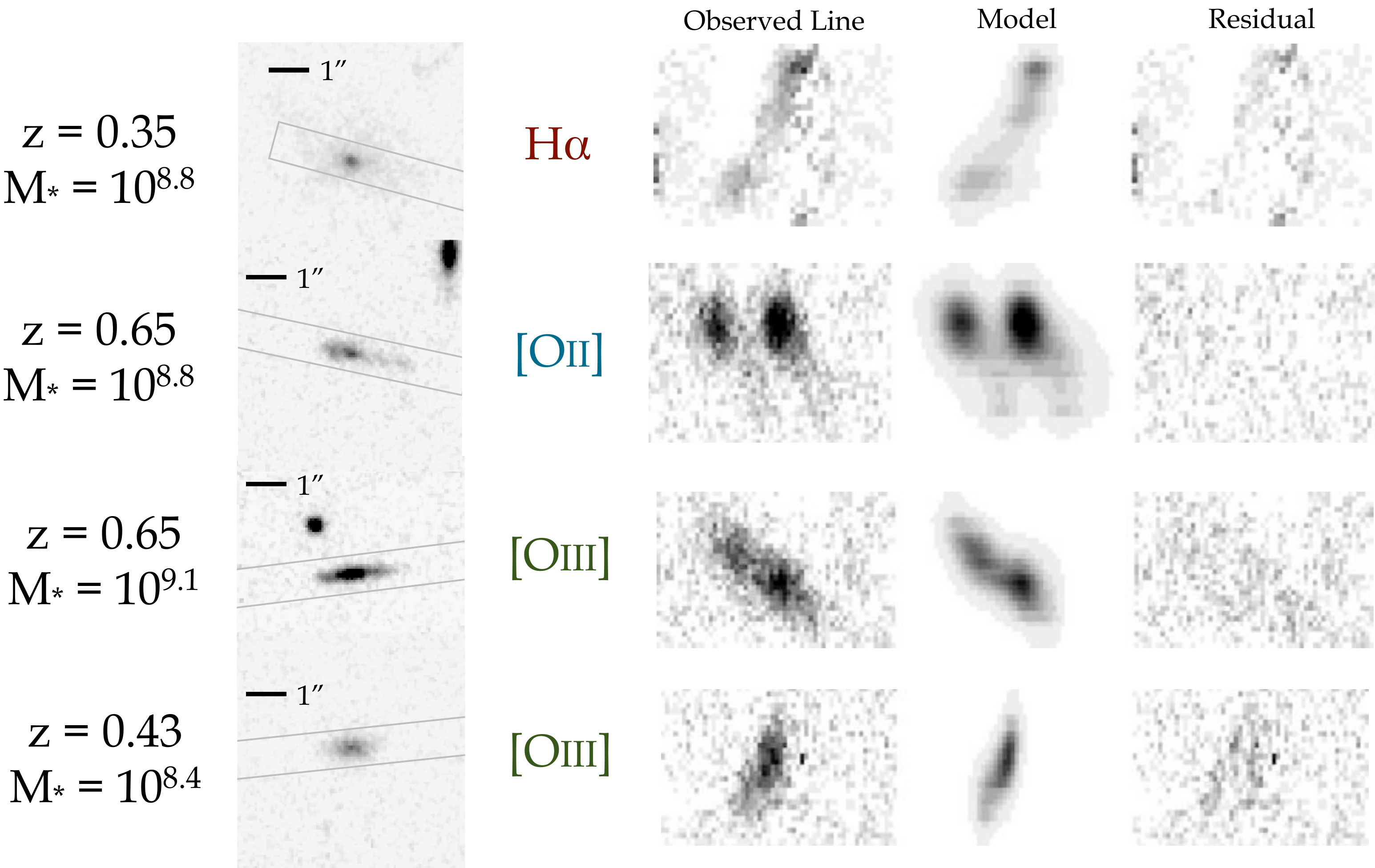}
\includegraphics[width=6.5in]{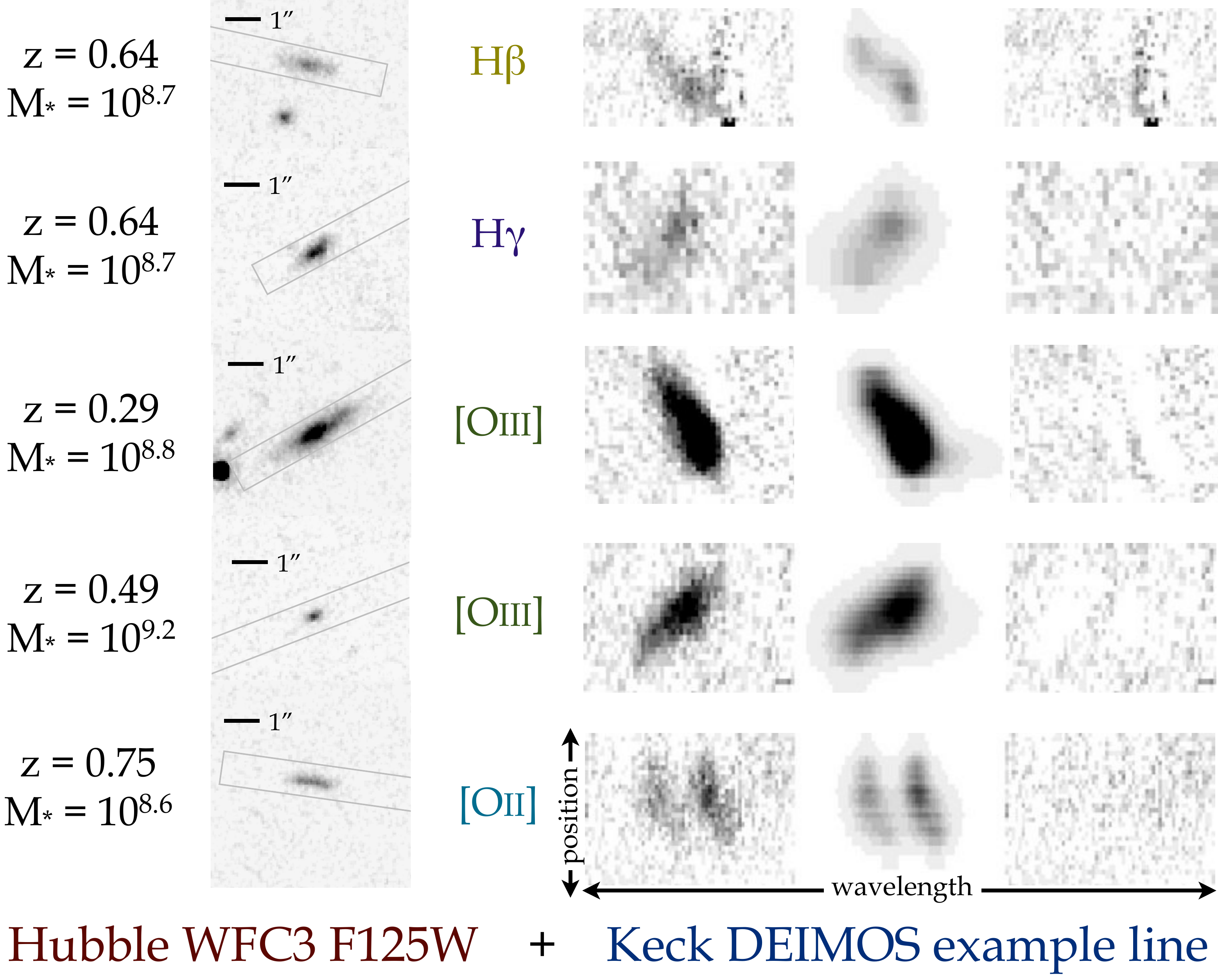}
\caption{Examples from the new intermediate redshift dwarf sample: {\bf (Left to right)}:  galaxy redshift and stellar mass; HST WFC3 F125W (H-band) image with the DEIMOS slitlet overlaid; a triptych of the 2-D spectrum, the modeled spectrum after blurring by both seeing and dispersion; the (data minus model) residual. Continued in Fig. \ref{fig:ex3}.}
\label{fig:ex1}
\end{figure*}

\subsection{Stellar Mass Estimates} \label{sec:stmass}

Stellar mass estimates were derived using the FAST (Fitting and Assessment of Synthetic Templates) code \citep{kriek2009} which fits stellar population synthesis templates \citep{bruzua2003} to broadband photometry adopting a \citet{chabri2003} initial mass function (IMF). For our photometric database \citep{newman2012}, we use $BVRiz$ from deep Subaru/XMM-Newton Deep Survey, SXDS \citep{furusa2008}, with mosaics prepared by \citet{cirasu2010}; $J$ (F125W) and $H$ (F160W) from CANDELS; and K photometry by the UKIDSS UDS Data Release 6 (DR6). We chose not to include photometry available at longer wavelengths due to blending confusion given the faintness of the sources targeted. We impose a floor of $\log{(\tau/}$yr) $=$ 8.5 for the timescale of the exponentially-declining star formation histories given this appears appropriate for star-forming galaxies \citep[e.g.,][]{wuyts2011}.  Sub-solar metallicities of $Z = $0.008, 0.004 as well as solar at 0.02 are permitted, consistent with the latest work on the metallicity of dwarf galaxies in the Local Group. FAST uses the Calzetti et al. (2000) reddening law.  In summary the stellar population grid limits are:
$\log{(\tau}$/yr) $=$ 8.5 - 10.0;
$\log{(}$age/yr) $=$ 7.0 - 10.1;
A${}_V$ $=$ 0.0 - 3.0;
metallicity $=$ 0.004-0.02.

We computed best-fitting model parameters with errors derived from Monte Carlo perturbations of the photometry, as well as the marginalized mean value and standard deviation of the parameter over the likelihood. We adopt the latter as they are more robust, however there are marginal differences for this particular dataset. 

It should be noted that the original set of stellar mass estimates used in the selection of the DEIMOS mask targets adopted a solar metallicity only, a Salpeter IMF, and the stellar population models of Charlot \& Bruzual (2007) which include more contribution from TP-ABG stars. Each of these choices is suited more to a higher mass sample than the dwarfs of this work. To compare to the final adopted estimates of this work, these alternate assumptions provided a systematic offset of $+$0.14 dex. For 7 galaxies at stellar masses $\log{(}$M${}_{\ast}$/M${}_{\odot})<$ 7.7 the results are particularly sensitive to these stellar population grid differences, but all estimates in the original set still remain below $\log{(}$M${}_{\ast}$/M${}_{\odot})<$ 8.7.

\begin{figure*}
\center
\includegraphics[width=6.6in]{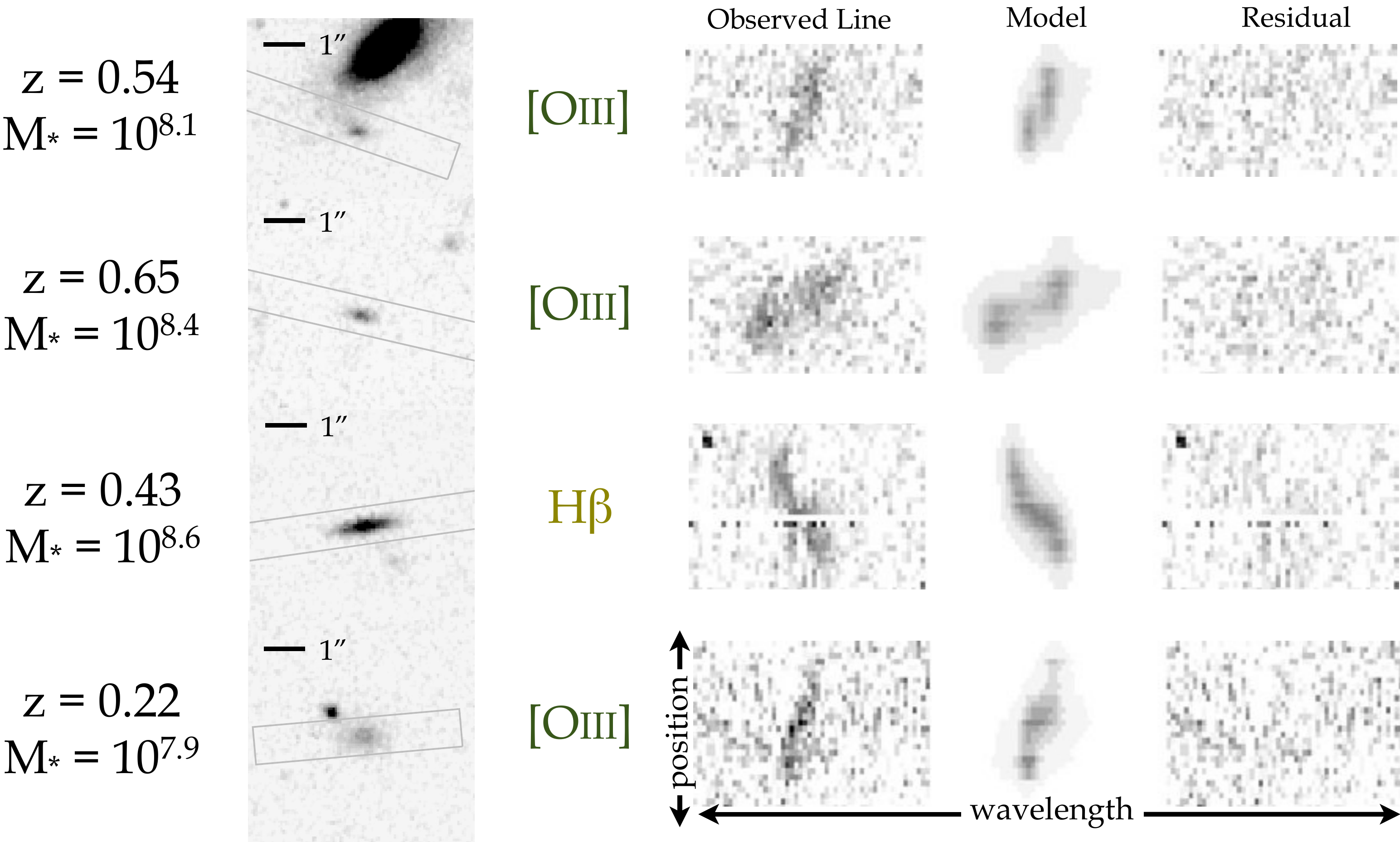}
\caption{As in Fig. \ref{fig:ex1}.}
\vspace{0.15in}
\label{fig:ex3}
\end{figure*}

\subsection{Image Measurements}

Examining the HST images of our sample, although most are small in angular extent (mean half-light radius, $\langle r_{1/2} \rangle \sim$ 1.2 kpc), the majority are regular in form with modest central concentrations. A high fraction have recognizable disks but few have prominent bulges.

We measure the radii and inclination of the dwarf galaxies in our sample assuming each galaxy is represented by a circular exponential disk. Scale radii and major-to-minor axis ratios are measured using GALFIT \citep{peng2010} with initial estimate distributions based on SExtractor \citep{bertin1996} in the same Monte Carlo method as described in previous work \citep{miller2011,miller2012a, miller2013}. We adopted this formalism for consistency with work at higher masses, and particularly to extract velocities at a consistent location, namely at 2.2 times the scale radius, $r_{2.2}$ \citep{miller2011}. The fiducial velocity measured at $r_{2.2}$ is denoted by $V_{2.2}$ and represents the peak rotational amplitude in a pure exponential disk. Although all of the dwarf galaxies of our sample are not necessarily best represented by a pure exponential, $V_{2.2}$ is generally well-suited in sampling the flattened portion of the rotation curve in galaxies \citep[see][for further discussion on this topic]{courte1997,miller2011}. For instance, allowing a free Sersic index fit in our sample produces a significantly more inconsistent set of effective radii by which to extract fiducial velocities for scaling relations.

While exponential disks are the simplest and most appropriate model for the morphologies of the majority of our dwarf sample, we also 
extract the velocity at a radius of 1 kpc (see Table \ref{table_of_results}).  These values are provided so that data samples analyzed with other techniques, namely predictions from numerical simulations, can more easily be compared to our values as exponential disk scale lengths via mock photometry measurements are not typically easily acquired.

\subsection{Dynamical Modeling}

The rotation curves for our sample were analyzed with the code developed in the previous studies of \citet{miller2011,miller2012a,miller2013}. We adopt the arctangent (arctan) functional form, and extract the velocity at 2.2 times the scale radius, $V_{2.2}$, correcting for the effect of inclination. This code has been described and tested extensively in our earlier papers and includes the following features. It adopts an asymmetric Gaussian for the emission line profile of each spatial bin, and accounts for both spatially-dependent velocity dispersion and surface-brightness profiles of the traced emission line in the spectrum. Including these features, the code also accounts for the effects of blurring by atmospheric seeing in the spatial direction, and blurring by instrumental dispersion in the spectral direction.  These are crucial ingredients for  recovering the intrinsic velocity profile of galaxies, particularly those with small angular extents. Examples of our 2-D emission line spectra and model fits are given in Figures \ref{fig:ex1} \& \ref{fig:ex3}. 

\smallskip

Table \ref{table_of_results} lists all of the galaxies in our sample, including each of the derived physical properties, where possible, which have been described in \S 3.


\section{Results}\label{sec:results}

\begin{figure*}
\center
\includegraphics[width=4.3in]{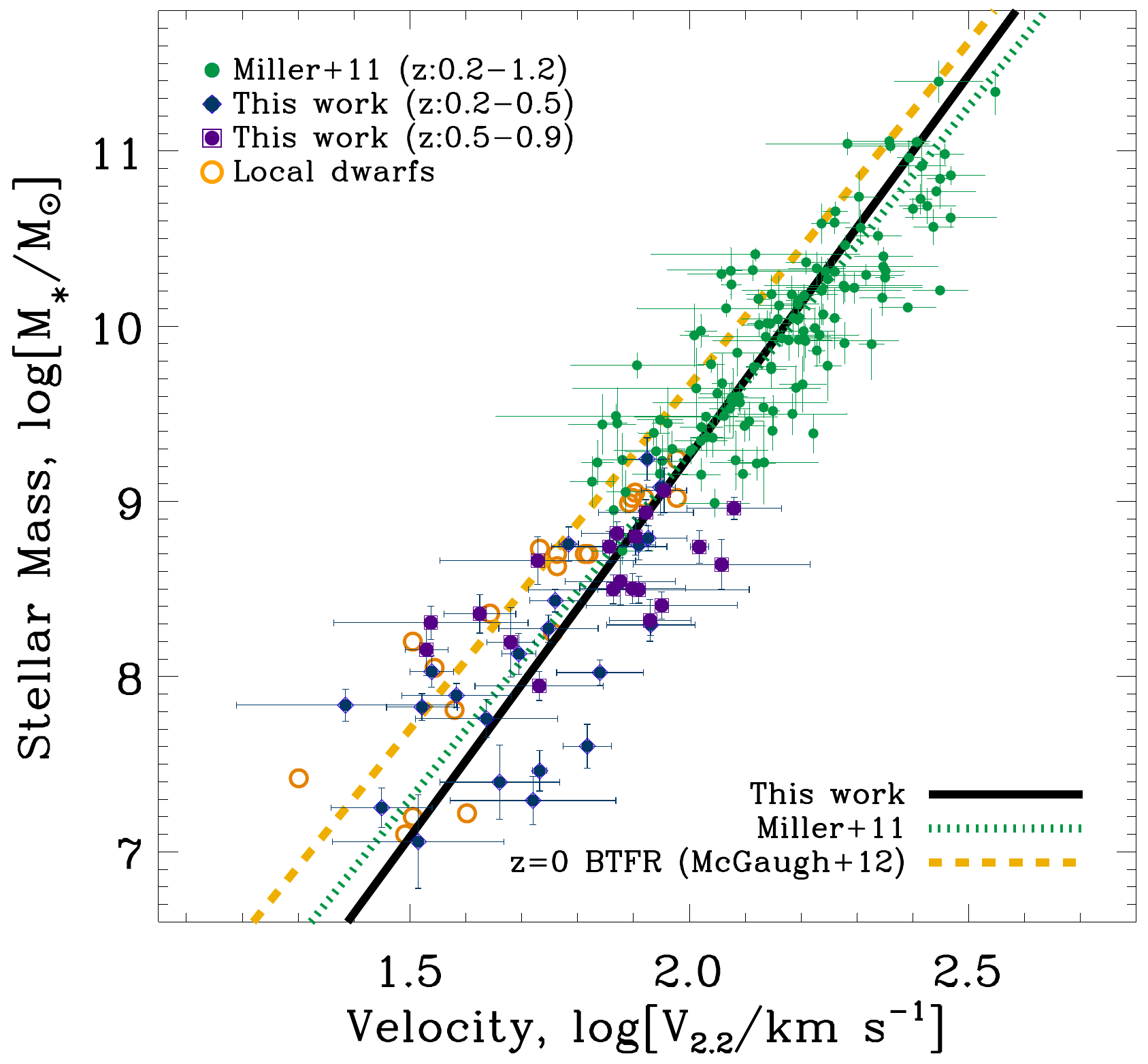}
\caption{The stellar mass Tully-Fisher (TF) relation: the new sample of low mass galaxies is shown according to two redshift intervals: blue diamonds ($0.2 < z < 0.5$) and purple squares ($0.5 < z < 0.9$). A more massive sample spanning the full redshift range is shown with green points \citep{miller2011} and the Local Group dwarf population at the targeted mass range, including the Magellanic Clouds, is plotted with golden rings. Lines represent best fits to the dwarf stellar mass TF relation (solid black), the intermediate redshift  sample \citep{miller2011} (dotted green), and the local Baryonic TF relation as determined by \citet{mcgaug2012} (yellow dashed).}
\vspace{0.2in}
\label{fig:smtf}
\end{figure*}

To evaluate the stellar-to-halo mass relation of our dwarf galaxy sample, we first consider the stellar mass Tully-Fisher (TF) relation (Fig. \ref{fig:smtf}), which represents the primary observational result of this paper,
alongside earlier work on higher masses at intermediate redshift \citep{miller2011}. For reference, we also indicate the Baryonic TF relation at $z\sim0$ \citep{mcgaug2012}, which includes the mass of the gas in the ordinate as well as the stellar mass. 

Our new data reveal a remarkably similar slope to the previously-measured Miller et al. relation at higher mass and similar redshifts (see Table \ref{table_smtfr} for the detailed comparison). Although aspects of the dwarf relation could suggest a somewhat steeper slope toward low mass than that inferred at high mass, this possible trend is marginal and requires more data for confirmation.  More interesting is the distribution of the locally-measured dwarfs relative to our sample at intermediate redshift.  As seen in Figure \ref{fig:smtf}, local galaxies lie mostly to the higher stellar mass (or rather slower velocity) side of the stellar mass TF relation. We will return to this topic in the next section.

\begin{figure}[h]
\center
\includegraphics[width=3.5in]{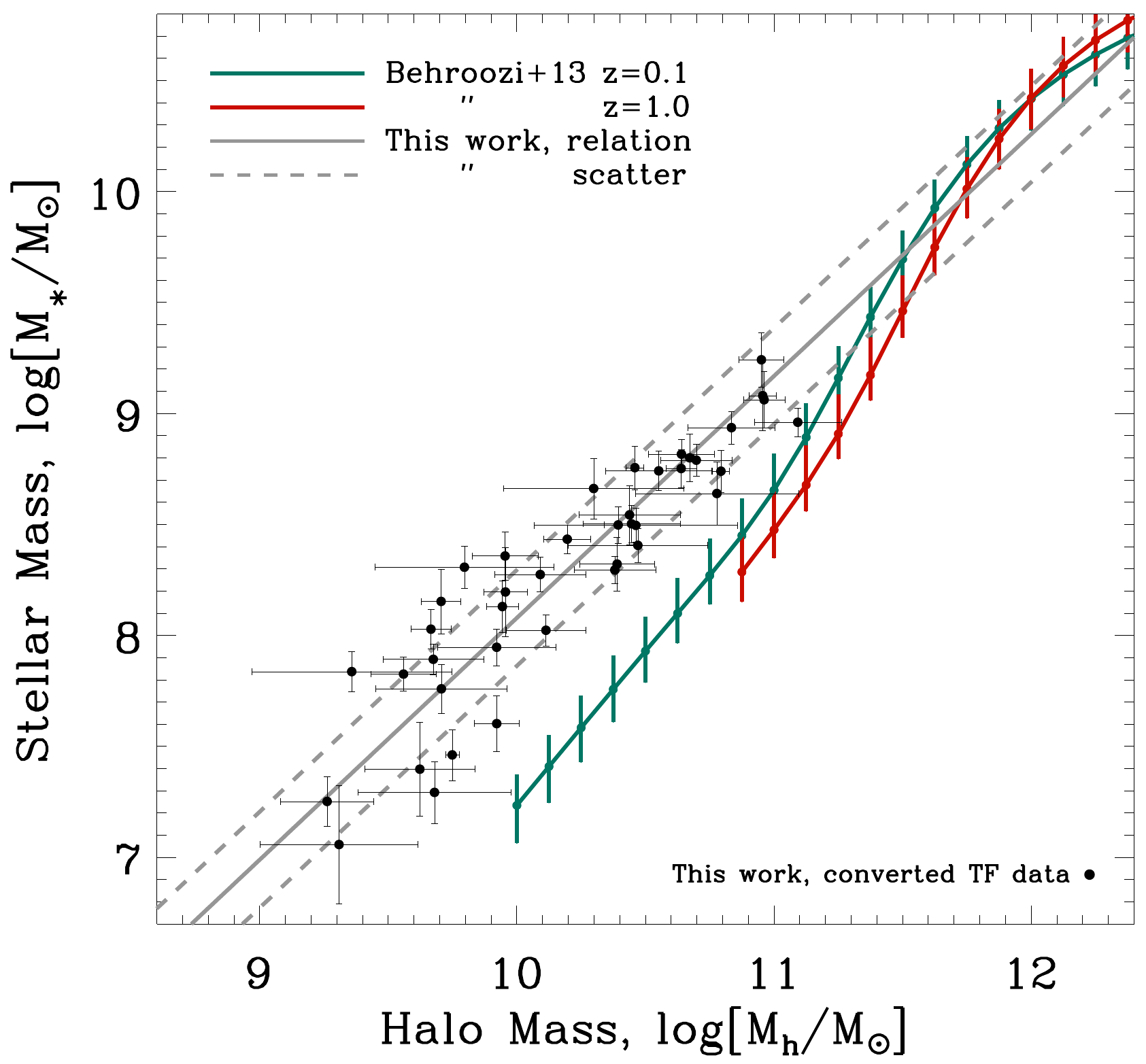}
\caption{The stellar-to-halo mass relation according to the abundance matching curves of \citet{behroo2013} (aqua and red curves at z=0.1 and z=1.0, respectively) derived in $\Lambda$CDM compared to that inferred for our sample of intermediate redshift dwarf galaxies by converting the measured $V_{2.2}$ velocities in the stellar mass Tully-Fisher relation to halo masses using the procedure discussed in \S3. Black points represent the data, the solid grey line is the best-fit relation and the scatter is indicated with dashed grey lines. }
\label{fig:shm}
\end{figure}

We next consider the relation between stellar mass and halo mass, by converting $V_{2.2}$ to $V_{200}$ in Fig. \ref{fig:shm} using the relation measured by \citet{reyes2012} with weak lensing. This relation is
effectively calibrated at log($M_{\ast}/M_{\odot}$) = 9.0, where $V_{2.2}/V_{200} = 1.05$ and has a slope of 0.53$\pm$0.03 (in terms of the relation of velocity ratio to stellar mass). To apply this in the 
low mass regime of interest here requires extrapolating this relationship to stellar masses of $10^7$ M${}_{\odot}$, which clearly introduces some uncertainty. We note that preliminary abundance matching results of the Next Generation Virgo Survey (Grossauer et al. in prep.) in the dwarf regime are consistent with the extrapolation of the stellar-to-halo mass relations from higher masses using abundance matching or weak lensing. Thus since the halo velocities in the Reyes et al. result are derived from weak-lensing as well, it is a fair hypothesis that the extrapolated relations are similar to what would be derived from the method conducted explicitly in the dwarf regime. 

Assuming spherical symmetry, we further convert $V_{200}$ velocities (where the over-density is 200 times the critical density, $200 \rho_{\mathrm{crit}} = M_{\rm 200} / \frac{4}{3} \pi R_{\rm 200}^3$) to halo masses via the standard adopted relation where $\log M_{200}/[h^{-1} M_{\odot}] = 3 \log{(G^{-1} V_{200} [\mathrm{km}~\mathrm{s}^{-1}])}$. The gravitational constant $G$ here is 4.3 $\times$ $10^{-6}$ kpc M${}_{\odot}^{-1}$ (km/s)${}^{2}$. While abundance matching results and calibrations based on the latest weak lensing studies \citep[e.g.,][]{leauth2012,behroo2013} produce consistent results when compared to the stellar mass TF relation above masses of $10^9$ M${}_{\odot}$, there is a clear divergence in the dwarf regime of these relations in Figure \ref{fig:shm}. We discuss the implications and uncertainties of this result in the next section.

\begin{small}
\begin{longtable*}{lllllllllll}
\caption{The Intermediate Redshift Dwarf Galaxy Sample} \label{table_of_results} \\
\hline \hline \\ [-1ex]
\multicolumn{1}{c}{R.A.} & 
\multicolumn{1}{c}{Dec.} &  
\multicolumn{1}{c}{$z$\tablenotemark{1}} &  
\multicolumn{1}{c}{sin($i$)\tablenotemark{2}} & 
\multicolumn{1}{c}{$r_s$\tablenotemark{3}[kpc]} & 
\multicolumn{1}{c}{log(M${}_{\ast}$/M${}_{\odot}$)\tablenotemark{4}} & 
\multicolumn{1}{c}{$V_{2.2}$\tablenotemark{5} [km s${}^{-1}$]}  & 
\multicolumn{1}{c}{$V_{1\mathrm{kpc}}$\tablenotemark{6}  [km s${}^{-1}$]}  & 
\multicolumn{1}{c}{log(M${}_{h}$/M${}_{\odot}$)\tablenotemark{7}} \\ [1ex] \hline \\
\endhead
\multicolumn{9}{c}{\textbf{Sample with resolved emission (N=41; ``Tully-Fisher sample''):}} \\ [1ex]
\hline \\  
34.287564 & -5.1435433 & 0.64 & 0.94 & 1.04$\pm$0.31 & 8.74$\pm$0.09 & 97.77$\pm$3.82 & 83.08$\pm$3.27 & 10.79$\pm$0.03 \\
34.303787 & -5.1458641 & 0.91 & 0.66 & 0.75$\pm$0.20 & 8.32$\pm$0.12 & 56.22$\pm$11.13 & 45.29$\pm$8.97 & 10.39$\pm$0.15 \\
34.358524 & -5.1515666 & 0.35 & 0.66 & 0.41$\pm$0.25 & 7.84$\pm$0.09 & 16.01$\pm$9.71 & 12.47$\pm$7.56 & 9.36$\pm$0.39 \\
34.398975 & -5.1525832 & 0.35 & 0.94 & 1.16$\pm$0.48 & 8.79$\pm$0.07 & 78.97$\pm$14.76 & 72.21$\pm$13.46 & 10.70$\pm$0.14 \\
34.289186 & -5.1563347 & 0.57 & 0.91 & 0.77$\pm$0.29 & 8.31$\pm$0.09 & 31.35$\pm$15.67 & 31.15$\pm$15.57 & 9.80$\pm$0.35 \\
34.359126 & -5.1571952 & 0.34 & 0.62 & 0.32$\pm$0.27 & 7.29$\pm$0.14 & 32.42$\pm$16.21 & 32.30$\pm$16.15 & 9.68$\pm$0.30 \\
34.336092 & -5.1585409 & 0.31 & 0.52 & 0.40$\pm$0.36 & 7.83$\pm$0.08 & 17.29$\pm$3.26 & 17.39$\pm$3.25 & 9.56$\pm$0.13 \\
34.489719 & -5.1596155 & 0.98 & 0.72 & 0.49$\pm$0.30 & 8.64$\pm$0.14 & 81.72$\pm$40.86 & 81.06$\pm$40.53 & 10.78$\pm$0.32 \\
34.404135 & -5.1638168 & 0.43 & 0.87 & 0.68$\pm$0.15 & 8.43$\pm$0.06 & 50.21$\pm$5.89 & 48.97$\pm$5.74 & 10.20$\pm$0.09 \\
34.310380 & -5.1642154 & 0.58 & 0.91 & 0.80$\pm$0.18 & 8.54$\pm$0.13 & 68.82$\pm$17.85 & 68.36$\pm$17.73 & 10.44$\pm$0.20 \\
34.367527 & -5.1671357 & 0.65 & 0.92 & 0.66$\pm$0.35 & 8.41$\pm$0.08 & 81.89$\pm$30.77 & 79.20$\pm$29.76 & 10.47$\pm$0.27 \\
34.440428 & -5.1676006 & 0.64 & 0.95 & 0.82$\pm$0.23 & 8.74$\pm$0.09 & 68.08$\pm$19.55 & 46.02$\pm$13.00 & 10.55$\pm$0.21 \\
34.473897 & -5.1691046 & 0.31 & 0.77 & 0.35$\pm$0.28 & 7.76$\pm$0.11 & 33.26$\pm$12.27 & 32.04$\pm$11.52 & 9.71$\pm$0.25 \\
34.446086 & -5.1700983 & 0.54 & 0.86 & 0.52$\pm$0.11 & 8.15$\pm$0.15 & 29.20$\pm$3.07 & 28.60$\pm$3.01 & 9.71$\pm$0.08 \\
34.279085 & -5.1710765 & 0.35 & 0.47 & 0.63$\pm$0.22 & 8.30$\pm$0.06 & 40.32$\pm$11.63 & 28.91$\pm$8.38 & 10.38$\pm$0.16 \\
34.350739 & -5.1712933 & 0.26 & 0.73 & 0.31$\pm$0.10 & 7.25$\pm$0.11 & 20.45$\pm$5.72 & 16.88$\pm$4.85 & 9.26$\pm$0.18 \\
34.503690 & -5.1738860 & 0.43 & 0.34 & 0.57$\pm$0.34 & 8.27$\pm$0.08 & 18.78$\pm$5.41 & 13.93$\pm$3.93 & 10.09$\pm$0.18 \\
34.432528 & -5.1760097 & 0.43 & 0.98 & 0.88$\pm$0.12 & 8.75$\pm$0.09 & 79.48$\pm$5.71 & 72.14$\pm$5.17 & 10.64$\pm$0.06 \\
34.466202 & -5.1780329 & 0.29 & 0.97 & 1.07$\pm$0.17 & 8.76$\pm$0.10 & 58.70$\pm$2.42 & 59.67$\pm$2.46 & 10.46$\pm$0.03 \\
34.429762 & -5.1785330 & 0.78 & 0.76 & 0.51$\pm$0.39 & 8.96$\pm$0.06 & 91.60$\pm$24.93 & 55.58$\pm$15.92 & 11.09$\pm$0.17 \\
34.434409 & -5.1793119 & 0.78 & 0.96 & 0.72$\pm$0.20 & 8.50$\pm$0.08 & 77.75$\pm$45.98 & 69.37$\pm$41.02 & 10.46$\pm$0.40 \\
34.517196 & -5.1793529 & 0.32 & 0.76 & 0.41$\pm$0.09 & 7.89$\pm$0.07 & 29.14$\pm$7.97 & 28.12$\pm$7.69 & 9.68$\pm$0.20 \\
34.250846 & -5.1797838 & 0.35 & 0.74 & 0.34$\pm$0.29 & 7.60$\pm$0.13 & 48.38$\pm$5.91 & 47.92$\pm$5.85 & 9.92$\pm$0.09 \\
34.415249 & -5.1820298 & 0.72 & 0.91 & 0.65$\pm$0.23 & 8.50$\pm$0.08 & 66.84$\pm$4.53 & 66.25$\pm$4.48 & 10.39$\pm$0.06 \\
34.296140 & -5.1827143 & 0.21 & 0.96 & 0.58$\pm$0.20 & 8.13$\pm$0.12 & 47.67$\pm$3.51 & 49.45$\pm$3.64 & 9.94$\pm$0.06 \\
34.362769 & -5.1829861 & 0.25 & 0.87 & 0.56$\pm$0.28 & 8.03$\pm$0.09 & 30.19$\pm$3.31 & 29.87$\pm$3.28 & 9.67$\pm$0.08 \\
34.477748 & -5.1834746 & 0.34 & 0.84 & 0.38$\pm$0.21 & 7.46$\pm$0.11 & 45.24$\pm$1.52 & 43.28$\pm$1.46 & 9.75$\pm$0.03 \\
34.423177 & -5.1861241 & 0.65 & 0.94 & 0.50$\pm$0.06 & 7.95$\pm$0.08 & 50.82$\pm$16.46 & 43.36$\pm$13.67 & 9.92$\pm$0.23 \\
34.282862 & -5.1864030 & 0.45 & 0.73 & 0.35$\pm$0.09 & 8.02$\pm$0.07 & 50.73$\pm$13.39 & 34.80$\pm$9.71 & 10.11$\pm$0.15 \\
34.520764 & -5.1871599 & 0.17 & 0.82 & 0.30$\pm$0.24 & 7.06$\pm$0.27 & 26.81$\pm$13.41 & 28.01$\pm$14.00 & 9.31$\pm$0.31 \\
34.413980 & -5.1882020 & 0.88 & 0.65 & 0.38$\pm$0.26 & 8.20$\pm$0.20 & 31.35$\pm$8.27 & 29.63$\pm$7.56 & 9.96$\pm$0.09 \\
34.523446 & -5.1899136 & 0.36 & 0.91 & 0.38$\pm$0.37 & 7.40$\pm$0.21 & 41.85$\pm$12.51 & 41.70$\pm$11.88 & 9.62$\pm$0.21 \\
34.254774 & -5.1914894 & 0.65 & 0.70 & 0.84$\pm$0.29 & 8.50$\pm$0.08 & 55.40$\pm$15.34 & 51.98$\pm$14.30 & 10.45$\pm$0.19 \\
34.359852 & -5.1916576 & 0.82 & 0.97 & 0.86$\pm$0.14 & 8.66$\pm$0.14 & 51.73$\pm$25.91 & 43.39$\pm$21.73 & 10.30$\pm$0.35 \\
34.417074 & -5.1950333 & 0.65 & 0.89 & 0.55$\pm$0.27 & 8.36$\pm$0.11 & 37.47$\pm$6.58 & 36.99$\pm$6.50 & 9.95$\pm$0.13 \\
34.370424 & -5.2035566 & 0.65 & 0.98 & 1.36$\pm$0.33 & 8.80$\pm$0.11 & 78.42$\pm$1.43 & 75.90$\pm$1.39 & 10.67$\pm$0.02 \\
34.465397 & -5.2073517 & 0.59 & 0.73 & 0.44$\pm$0.23 & 8.82$\pm$0.07 & 54.11$\pm$14.78 & 50.42$\pm$13.23 & 10.64$\pm$0.13 \\
34.446346 & -5.1498754 & 0.49 & 0.94 & 0.92$\pm$0.29 & 9.24$\pm$0.12 & 79.21$\pm$8.48 & 78.91$\pm$7.87 & 10.95$\pm$0.09 \\
34.283345 & -5.1519078 & 0.65 & 0.98 & 1.08$\pm$0.13 & 9.06$\pm$0.13 & 87.92$\pm$8.62 & 82.34$\pm$7.73 & 10.96$\pm$0.08 \\
34.392003 & -5.1590678 & 0.56 & 0.92 & 0.92$\pm$0.09 & 8.94$\pm$0.07 & 77.18$\pm$17.10 & 65.16$\pm$14.63 & 10.83$\pm$0.17 \\
34.275351 & -5.1724717 & 0.41 & 0.88 & 0.74$\pm$0.44 & 9.08$\pm$0.16 & 77.91$\pm$5.11 & 76.85$\pm$5.05 & 10.96$\pm$0.05 \\ [1ex]
\hline \\ 
\multicolumn{9}{l}{\textbf{Sample with unresolved emission (N=18)}} \\ [1ex]
\hline \\ 
34.452142 & -5.1527207 & 0.62 & - & 0.57$\pm$0.13 & 8.29$\pm$0.10 & - & - & - \\
34.490433 & -5.1591549 & 0.29 & - & 0.62$\pm$0.47 & 8.58$\pm$0.09 & - & - & - \\
34.333114 & -5.1617036 & 0.55 & - & 0.34$\pm$0.20 & 8.45$\pm$0.13 & - & - & - \\
34.501182 & -5.1717299 & 0.57 & - & 0.76$\pm$0.44 & 8.88$\pm$0.12 & - & - & - \\
34.269635 & -5.1725942 & 0.52 & - & 0.91$\pm$0.07 & 8.36$\pm$0.08 & - & - & - \\
34.321552 & -5.1734201 & 0.65 & - & 0.50$\pm$0.14 & 8.45$\pm$0.08 & - & - & - \\
34.393462 & -5.1748474 & 0.43 & - & 0.58$\pm$0.10 & 8.33$\pm$0.08 & - & - & - \\
34.338840 & -5.1827413 & 0.43 & - & 0.79$\pm$0.34 & 8.83$\pm$0.07 & - & - & - \\
34.344560 & -5.1913127 & 0.39 & - & 0.29$\pm$0.12 & 7.95$\pm$0.11 & - & - & - \\
34.496066 & -5.2060556 & 0.80 & - & 0.67$\pm$0.13 & 8.72$\pm$0.17 & - & - & - \\
34.266036 & -5.1594535 & 0.82 & - & 0.81$\pm$0.27 & 8.35$\pm$0.16 & - & - & - \\
34.338844 & -5.1653685 & 0.79 & - & 0.39$\pm$0.34 & 8.50$\pm$0.10 & - & - & - \\
34.481657 & -5.1674153 & 0.75 & - & 0.67$\pm$0.23 & 8.66$\pm$0.06 & - & - & - \\
34.370433 & -5.1823730 & 0.20 & - & 0.22$\pm$0.15 & 7.57$\pm$0.16 & - & - & - \\
34.481431 & -5.1833789 & 0.46 & - & 0.67$\pm$0.26 & 8.95$\pm$0.09 & - & - & - \\
34.514295 & -5.1839202 & 0.39 & - & 0.43$\pm$0.19 & 8.23$\pm$0.09 & - & - & - \\
34.341006 & -5.1872395 & 0.44 & - & 0.27$\pm$0.24 & 8.10$\pm$0.08 & - & - & - \\
34.375929 & -5.1970098 & 0.60 & - & 0.80$\pm$0.07 & 8.38$\pm$0.16 & - & - & - \\
[1ex]
\hline \\ 
\multicolumn{9}{l}{\textbf{Sample with no emission detected (N=23)}} \\ [1ex]
\hline \\ 
34.295579 & -5.1471624 & (0.61) & - & 0.53$\pm$0.11 & 8.73$\pm$0.12 & - & - & - \\
34.403433 & -5.1479603 & (0.38) & - & 0.34$\pm$0.10 & 7.89$\pm$0.10 & - & - & - \\
34.273705 & -5.1537420 & (0.29) & - & 0.50$\pm$0.22 & 8.07$\pm$0.10 & - & - & - \\
34.422943 & -5.1545570 & (0.58) & - & 0.73$\pm$0.26 & 8.33$\pm$0.16 & - & - & - \\
34.354658 & -5.1553817 & (0.43) & - & 0.27$\pm$0.22 & 8.20$\pm$0.07 & - & - & - \\
34.396846 & -5.1563103 & (0.43) & - & 0.48$\pm$0.06 & 7.95$\pm$0.09 & - & - & - \\
34.410996 & -5.1571028 & (0.32) & - & 0.34$\pm$0.28 & 7.85$\pm$0.09 & - & - & - \\
34.327556 & -5.1592369 & (0.22) & - & 0.25$\pm$0.20 & 7.25$\pm$0.25 & - & - & - \\
34.454441 & -5.1590304 & (0.42) & - & 0.25$\pm$0.20 & 7.90$\pm$0.27 & - & - & - \\
34.509970 & -5.1648048 & (0.66) & - & 0.36$\pm$0.11 & 8.37$\pm$0.14 & - & - & - \\
34.507877 & -5.1696592 & (0.75) & - & 0.84$\pm$0.21 & 8.62$\pm$0.15 & - & - & - \\
34.305389 & -5.1736575 & (0.28) & - & 0.38$\pm$0.09 & 7.65$\pm$0.16 & - & - & - \\
34.313890 & -5.1745230 & (0.32) & - & 0.39$\pm$0.26 & 7.81$\pm$0.09 & - & - & - \\
34.398413 & -5.1746398 & (0.76) & - & 0.59$\pm$0.43 & 8.51$\pm$0.13 & - & - & - \\
34.258658 & -5.1763573 & (0.66) & - & 0.63$\pm$0.06 & 8.73$\pm$0.10 & - & - & - \\
34.447737 & -5.1786612 & (0.52) & - & 0.46$\pm$0.09 & 8.10$\pm$0.16 & - & - & - \\
34.474809 & -5.1845825 & (0.31) & - & 0.43$\pm$0.38 & 7.86$\pm$0.11 & - & - & - \\
34.495861 & -5.1862426 & (0.27) & - & 0.26$\pm$0.09 & 7.10$\pm$0.22 & - & - & - \\
34.468164 & -5.1934894 & (0.59) & - & 0.50$\pm$0.13 & 8.05$\pm$0.07 & - & - & - \\
34.453041 & -5.1942366 & (0.39) & - & 0.39$\pm$0.13 & 7.46$\pm$0.20 & - & - & - \\
34.438733 & -5.1996580 & (0.28) & - & 0.25$\pm$0.24 & 8.13$\pm$0.11 & - & - & - \\
34.507926 & -5.2060951 & (0.64) & - & 0.78$\pm$0.50 & 8.47$\pm$0.22 & - & - & - \\
34.372178 & -5.2088495 & (0.31) & - & 0.36$\pm$0.25 & 7.88$\pm$0.12 & - & - & - \\
[1ex]
\hline \\
\caption{[1] Redshifts are spectroscopic except when in parentheses (photometric); [2] Sine of inclination; [3] Exponential scale radius (kpc); [4] Stellar mass in log(M${}_{\ast}$/M${}_{\odot}$); [5] Rotation velocity $V_{2.2}$ km s${}^{-1}$; [6] Rotation velocity at 1 kpc (km s${}^{-1}$); [7] Halo mass inferred as described in \S 3, log(M${}_h$/M${}_{\odot}$).}
\end{longtable*}
\end{small}

\begin{small}
\begin{longtable*}{lllllllllll}
\caption{Stellar mass TF relation fits} \label{table_smtfr} \\
 \hline \hline \\ [-1ex]
\multicolumn{1}{c}{$z$ range} & 
\multicolumn{1}{c}{$\langle z \rangle$} &  
\multicolumn{1}{c}{N} &  
\multicolumn{1}{c}{$a$\tablenotemark{1}} & 
\multicolumn{1}{c}{$b$\tablenotemark{2}} & 
\multicolumn{1}{c}{$\sigma_{int,V}$\tablenotemark{3}} & 
\multicolumn{1}{c}{med $\sigma_{V}$\tablenotemark{4}}  & 
\multicolumn{1}{c}{$rms_V$\tablenotemark{5}}  & 
\multicolumn{1}{c}{$\sigma_{int,M}$\tablenotemark{6}}  & 
\multicolumn{1}{c}{med $\sigma_{M}$\tablenotemark{7}}  & 
\multicolumn{1}{c}{$\mathrm{rms}_M$\tablenotemark{8}} \\ [1ex] \hline \\
\endhead
\multicolumn{11}{l}{\textbf{M${}_{\ast} > 10^9$ M${}_{\odot}$ Relation (Miller et al. 2011):  $M_{\ast}$  vs.  $V$($r_{2.2}$):}} \\ [1ex]
\hline \\  
{0.2$<$z$\le$1.3} & {0.64} & {129} & {1.718$\pm$0.415} & {3.869$\pm$0.193} & {0.058} & {0.022} & {0.083} & {0.224} & {0.091} & {0.323} \\[1ex] \hline \\ 
\multicolumn{11}{l}{\textbf{Dwarf Mass Relation (this work):  $M_{\ast}$ vs.  $V$($r_{2.2}$)}} \\ [1ex]
\hline \\ 
{0.2$<$z$\le$1.0} & {0.64} & {41} & {0.566$\pm$0.476} & {4.347$\pm$0.623} & {0.064} & {0.078} & {0.122} & {0.375} & {0.095} & {0.532} \\ [1ex]
\hline \\
\caption{[1] y-int in $M_{\ast}/M_{\odot}$ dex assuming scatter in $V$/km s$^{-1}$ dex ; [2] slope assuming scatter in $V$/km s$^{-1}$ dex; [3] internal scatter in $V$/km s$^{-1}$ dex; [4] median velocity error in $V$/km s$^{-1}$ dex; [5] total scatter in $V$/km s$^{-1}$ dex; [6] internal scatter in $M_{\ast}/M_{\odot}$ dex; [7] median stellar mass error in $M_{\ast}/M_{\odot}$ dex; [8] total scatter in $M_{\ast}/M_{\odot}$ dex.}
\end{longtable*}
\end{small}

\section{Discussion} \label{disc}

\subsection{Is the Local Group Population Representative?}

Figure \ref{fig:smtf} shows a striking effect, namely that local galaxies appear to be drawn from a particular subset of the stellar mass TF relation defined by our larger sample. It seems unlikely that this is an evolutionary effect (relating for example to changes in the dark matter density profiles or effect of baryonic feedback) given there is no obvious trend with redshift within our sample itself. Otherwise this would result in the curious conclusion that the relation between baryons and dark matter becomes tighter over time when non-linear exchanges dominate. A more likely explanation is that we are sensitive to sources at intermediate redshift that are not typically found locally. For example, low mass galaxies generally may have a wider range of gas fractions which could explain the direction of the offset and increased scatter. Clearly we are biased in our survey towards dwarf galaxies with higher gas fractions that allow the tracing of extended emission lines for our kinematic analysis. Also the field of selection within the DEIMOS mask samples on average lower density environments compared to the group environment locally. This may proportionally affect dwarfs more than higher mass galaxies due to a satellite effect: gas is lost to interactions with other satellites and the host galaxy, whereby in the field and in voids dwarf galaxies are more likely to keep their gas.

Taking this argument further, it is interesting to speculate what role gas fractions may play in reconciling the abundance matching discrepancy in Fig. \ref{fig:shm}. The rank order for matching haloes to observed mass functions could vary widely moving into the dwarf regime with the inclusion of gas mass. Separately, could including gas and realistic feedback prescriptions in the simulations used to construct halo trees for abundance matching have an appreciable effect towards reconciling the offset? Neither of the stellar-to-halo mass relations being compared for consistency in Fig. \ref{fig:shm} take full account of the effect of gas. Unfortunately, future progress on this front will be hindered by the difficulty in measuring gas masses in the dwarf regime as well as at higher-$z$ where they appear to be greater on average than at low redshift \citep{taccon2010}.  

\subsection{The Role of Feedback? New Dark Matter Physics?}

Implementing episodic supernova feedback in hydrodynamical simulations \citep{govern2012} has been successful in flattening dark matter density profiles of haloes over time. This may even resolve a related, long-standing difficulty whereby simulations predicting inner density profiles that are too ``cuspy'' \citep{pontze2012}. Although this effect can be reproduced in both adaptive mesh refinement simulations \citep{teyssi2013}, and smoothed particle hydrodynamics, other workers have had difficulty reproducing the result in alternative prescriptions of the same feedback process. While this action may work to reconcile the offset present in Fig. \ref{fig:shm} if the redistribution of matter was efficient out to 2.2 kpc, no amount of feedback has been effective at flattening inner density profiles below stellar masses of $10^7 $M${}_{\odot}$ \citep{govern2012}. Thus if the offset observed in Fig. \ref{fig:smtf} were to continue $\sim$1 dex lower in stellar mass, then this feedback mechanism could not alone reconcile the matched and observed relations. Additionally there are arguments that the energy of the necessary feedback for this effect exceeds what would likely occur over the typical star-formation history of most dwarfs \citep{garris2013}. 

Alternatively, a combination of early feedback effects with ram pressure stripping and tidal heating by both the host halo and disk appear to extract enough energy from the gravitational potential of the host to reproduce the observations in some examples \citep{arraki2012,zoloto2012,brooks2013}. In future work we hope to better ascertain how environmental effects may be responsible for our observed trends by determining whether targeted dwarfs are isolated or likely satellites. If most of the dwarfs are isolated, as suspected given their inferred gas-rich state, then it will be difficult to use this mechanism to explain Fig. \ref{fig:shm}.

Ultimately a more fundamental adjustment to the dark matter model could be required \citep{weinbe2013}. Alternate dark matter models that were not fully explored before the last decade of consensus around CDM include warm dark matter models \citep[e.g.,][]{benson2012,lovell2012},  various self-interacting dark matter models \citep{sperge2000,rocha2013,peter2013}, and flavor-mixed dark matter \citep{medved2000,medved2012}. Any of these may potentially hold the key in resolving observed tensions in the CDM predictions for galaxy formation and evolution, however work has only just begun exploring these alternative models in detail.

\subsection{The Utility of the Stellar to Halo Mass Relation}

Figure \ref{fig:shm} is difficult to interpret without acknowledging CDM simulations generate too much substructure. When galaxies are matched to halos from the observed stellar mass functions, they are placed in halos which are too large to be consistent with the observed fiducial rotation velocities related to their stellar mass. This discrepancy increases as the (mis-)matching propagates into the dwarf galaxy regime.

We recognize that the utility of Fig. \ref{fig:shm} in the dwarf regime relies on certain aspects of the CDM halo paradigm for us to convert observed velocities at observable radii to those of the unobservable halo. Several effects could explain the discrepancy in Fig. \ref{fig:shm}, the most obvious of which is that the linear extrapolation of the \citet{reyes2012} relation (connecting $V_{2.2}$ to $V_{200}$) to the dwarf regime could be inaccurate, as discussed in \S 4. The stellar mass estimates could be incorrect, if for example the stellar population synthesis models used are inappropriate for low mass galaxies.\footnote{Recent work has suggested that low mass galaxy stellar mass estimates are likely underestimated when using exponential star-formation histories (Dominguez et al., in prep.), exacerbating the discrepancy in Fig. \ref{fig:shm}.}. Most importantly, the gas mass should be accounted for to create a `Baryonic' TF consistency check; this might reconcile the offset seen in Figure \ref{fig:smtf} and affect the abundance matching method currently based only on the stellar mass distribution. Finally, only when such caveats can be laid to rest might we seriously consider powerful feedback effects which would modify dark matter density profiles out to 2.2 kpc in cosmological simulations or dark matter physics beyond the standard $\Lambda$CDM paradigm.

\subsection{A More Direct Observational Test}

Noting the considerable uncertainties in the interpretation of Figure \ref{fig:shm}, a more direct test would be to extract quantities such as $V_{2.2}$ from halos within the cosmological simulations, in order to conduct a comparison in the observed space, i.e. Figure \ref{fig:smtf}.  Ideally the abundance matching curve with its scatter would not be extracted in terms of the stellar mass Tully-Fisher relation, but rather the full baryonic Tully-Fisher relation with the inclusion of gas mass. Not accounting for the effects of gas in a comprehensive manner may ultimately lead to the mismatch we see between simulated CDM predictions and observations in each context:  from early versions of the ``missing satellite problem'', to the ``Too Big to Fail'' framing within the Milky Way, to the semi-empirical curves of abundance matching diverging away from the featureless TF relations. A more careful comparison, one which includes gas rather than stellar mass alone, is needed to explore these ideas further and will be addressed in future work.

\acknowledgments

SHM thanks the Rhodes Trust, the British Federation of Women Graduates, the sub-department of Astrophysics and New College at the University of Oxford, the University of California, Riverside, and the California Institute of Technology for supporting her work. The spectroscopic data was secured with the W.M. Keck Observatory on Mauna Kea. We thank the observatory staff for their dedication and support. The authors recognize and acknowledge the cultural role and reverence that the summit of Mauna Kea has always had with the indigenous Hawaiian community, and we are most fortunate to have the opportunity to conduct observations from this mountain.

{\it Facilities:} \facility{Keck II (DEIMOS)}, \facility{HST (WFC3/IR)}.


\bibliographystyle{apj}

\bibliography{mybib}

\begin{thebibliography}{}

\bibitem[\protect\citeauthoryear{{Arraki} et~al.}{{Arraki}
  et~al.}{2012}]{arraki2012}
{Arraki}, K.~S., {Klypin}, A., {More}, S.,  \& {Trujillo-Gomez}, S. 2012, ArXiv
  e-prints

\bibitem[\protect\citeauthoryear{{Behroozi}, {Conroy}, \&
  {Wechsler}}{{Behroozi} et~al.}{2010}]{behroo2010}
{Behroozi}, P.~S., {Conroy}, C.,  \& {Wechsler}, R.~H. 2010, \apj, 717, 379

\bibitem[\protect\citeauthoryear{{Behroozi}, {Wechsler}, \&
  {Conroy}}{{Behroozi} et~al.}{2013}]{behroo2013}
{Behroozi}, P.~S., {Wechsler}, R.~H.,  \& {Conroy}, C. 2013, \apj, 770, 57

\bibitem[\protect\citeauthoryear{{Benson}}{{Benson}}{2012}]{benson2012}
{Benson}, A. 2012, New Astronomy, 17, 175

\bibitem[\protect\citeauthoryear{{Bertin} \& {Arnouts}}{{Bertin} \&
  {Arnouts}}{1996}]{bertin1996}
{Bertin}, E.,  \& {Arnouts}, S. 1996, \aaps, 117, 393

\bibitem[\protect\citeauthoryear{{Boylan-Kolchin}, {Bullock}, \&
  {Kaplinghat}}{{Boylan-Kolchin} et~al.}{2011}]{boylan2011}
{Boylan-Kolchin}, M., {Bullock}, J.~S.,  \& {Kaplinghat}, M. 2011, \mnras, 415,
  L40

\bibitem[\protect\citeauthoryear{{Brammer}, {van Dokkum}, \& {Coppi}}{{Brammer}
  et~al.}{2008}]{bramme2008}
{Brammer}, G.~B., {van Dokkum}, P.~G.,  \& {Coppi}, P. 2008, \apj, 686, 1503

\bibitem[\protect\citeauthoryear{{Brooks} et~al.}{{Brooks}
  et~al.}{2013}]{brooks2013}
{Brooks}, A.~M., {Kuhlen}, M., {Zolotov}, A.,  \& {Hooper}, D. 2013, \apj, 765,
  22

\bibitem[\protect\citeauthoryear{{Bruzual} \& {Charlot}}{{Bruzual} \&
  {Charlot}}{2003}]{bruzua2003}
{Bruzual}, G.,  \& {Charlot}, S. 2003, \mnras, 344, 1000

\bibitem[\protect\citeauthoryear{{Bullock} et~al.}{{Bullock}
  et~al.}{2001}]{bulloc2001}
{Bullock}, J.~S., {Dekel}, A., {Kolatt}, T.~S., {Kravtsov}, A.~V., {Klypin},
  A.~A., {Porciani}, C.,  \& {Primack}, J.~R. 2001, \apj, 555, 240

\bibitem[\protect\citeauthoryear{{Bullock}, {Kravtsov}, \&
  {Weinberg}}{{Bullock} et~al.}{2000}]{bulloc2000}
{Bullock}, J.~S., {Kravtsov}, A.~V.,  \& {Weinberg}, D.~H. 2000, \apj, 539, 517

\bibitem[\protect\citeauthoryear{{Chabrier}}{{Chabrier}}{2003}]{chabri2003}
{Chabrier}, G. 2003, \pasp, 115, 763

\bibitem[\protect\citeauthoryear{{Cirasuolo} et~al.}{{Cirasuolo}
  et~al.}{2010}]{cirasu2010}
{Cirasuolo}, M., {McLure}, R.~J., {Dunlop}, J.~S., {Almaini}, O., {Foucaud},
  S.,  \& {Simpson}, C. 2010, \mnras, 401, 1166

\bibitem[\protect\citeauthoryear{{Conroy} et~al.}{{Conroy}
  et~al.}{2007}]{conroy2007}
{Conroy}, C., et~al. 2007, \apj, 654, 153

\bibitem[\protect\citeauthoryear{{Courteau}}{{Courteau}}{1997}]{courte1997}
{Courteau}, S. 1997, \aj, 114, 2402

\bibitem[\protect\citeauthoryear{{Diemand}, {Kuhlen}, \& {Madau}}{{Diemand}
  et~al.}{2007}]{dieman2007}
{Diemand}, J., {Kuhlen}, M.,  \& {Madau}, P. 2007, \apj, 667, 859

\bibitem[\protect\citeauthoryear{{Diemand} et~al.}{{Diemand}
  et~al.}{2008}]{dieman2008}
{Diemand}, J., {Kuhlen}, M., {Madau}, P., {Zemp}, M., {Moore}, B., {Potter},
  D.,  \& {Stadel}, J. 2008, \nat, 454, 735

\bibitem[\protect\citeauthoryear{{Font} et~al.}{{Font} et~al.}{2011}]{font2011}
{Font}, A.~S., et~al. 2011, \mnras, 417, 1260

\bibitem[\protect\citeauthoryear{{Furusawa} et~al.}{{Furusawa}
  et~al.}{2008}]{furusa2008}
{Furusawa}, H., {Kosugi}, G., {Akiyama}, M., {Takata}, T., {Sekiguchi}, K.,  \&
  {Furusawa}, J. 2008, in Astronomical Society of the Pacific Conference
  Series, Vol. 399, Panoramic Views of Galaxy Formation and Evolution, ed.
  T.~{Kodama}, T.~{Yamada}, \& K.~{Aoki}, 131

\bibitem[\protect\citeauthoryear{{Garrison-Kimmel} et~al.}{{Garrison-Kimmel}
  et~al.}{2013}]{garris2013}
{Garrison-Kimmel}, S., {Rocha}, M., {Boylan-Kolchin}, M., {Bullock}, J.~S.,  \&
  {Lally}, J. 2013, \mnras, 433, 3539

\bibitem[\protect\citeauthoryear{{Geha} et~al.}{{Geha} et~al.}{2006}]{geha2006}
{Geha}, M., {Blanton}, M.~R., {Masjedi}, M.,  \& {West}, A.~A. 2006, \apj, 653,
  240

\bibitem[\protect\citeauthoryear{{Governato} et~al.}{{Governato}
  et~al.}{2012}]{govern2012}
{Governato}, F., et~al. 2012, \mnras, 422, 1231

\bibitem[\protect\citeauthoryear{{Grogin} et~al.}{{Grogin}
  et~al.}{2011}]{grogin2011}
{Grogin}, N.~A., et~al. 2011, \apjs, 197, 35

\bibitem[\protect\citeauthoryear{{Klypin} et~al.}{{Klypin}
  et~al.}{1999}]{klypin1999}
{Klypin}, A., {Kravtsov}, A.~V., {Valenzuela}, O.,  \& {Prada}, F. 1999, \apj,
  522, 82

\bibitem[\protect\citeauthoryear{{Koekemoer} et~al.}{{Koekemoer}
  et~al.}{2011}]{koekem2011}
{Koekemoer}, A.~M., et~al. 2011, \apjs, 197, 36

\bibitem[\protect\citeauthoryear{{Kriek} et~al.}{{Kriek}
  et~al.}{2009}]{kriek2009}
{Kriek}, M., {van Dokkum}, P.~G., {Labb{\'e}}, I., {Franx}, M., {Illingworth},
  G.~D., {Marchesini}, D.,  \& {Quadri}, R.~F. 2009, \apj, 700, 221

\bibitem[\protect\citeauthoryear{{Leauthaud} et~al.}{{Leauthaud}
  et~al.}{2012}]{leauth2012}
{Leauthaud}, A., et~al. 2012, \apj, 744, 159

\bibitem[\protect\citeauthoryear{{Lovell} et~al.}{{Lovell}
  et~al.}{2012}]{lovell2012}
{Lovell}, M.~R., et~al. 2012, \mnras, 420, 2318

\bibitem[\protect\citeauthoryear{{Mandelbaum} et~al.}{{Mandelbaum}
  et~al.}{2006}]{mandel2006}
{Mandelbaum}, R., {Seljak}, U., {Kauffmann}, G., {Hirata}, C.~M.,  \&
  {Brinkmann}, J. 2006, \mnras, 368, 715

\bibitem[\protect\citeauthoryear{{McGaugh}}{{McGaugh}}{2012}]{mcgaug2012}
{McGaugh}, S.~S. 2012, \aj, 143, 40

\bibitem[\protect\citeauthoryear{{Medvedev}}{{Medvedev}}{2000}]{medved2000}
{Medvedev}, M.~V. 2000, ArXiv Astrophysics e-prints

\bibitem[\protect\citeauthoryear{{Medvedev}}{{Medvedev}}{2012}]{medved2012}
{Medvedev}, M.~V. 2012, in APS April Meeting Abstracts, G7007

\bibitem[\protect\citeauthoryear{{Miller} et~al.}{{Miller}
  et~al.}{2011}]{miller2011}
{Miller}, S.~H., {Bundy}, K., {Sullivan}, M., {Ellis}, R.~S.,  \& {Treu}, T.
  2011, \apj, 741, 115

\bibitem[\protect\citeauthoryear{{Miller} et~al.}{{Miller}
  et~al.}{2012}]{miller2012a}
{Miller}, S.~H., {Ellis}, R.~S., {Sullivan}, M., {Bundy}, K., {Newman}, A.~B.,
  \& {Treu}, T. 2012, \apj, 753, 74

\bibitem[\protect\citeauthoryear{{Miller}, {Sullivan}, \& {Ellis}}{{Miller}
  et~al.}{2013}]{miller2013}
{Miller}, S.~H., {Sullivan}, M.,  \& {Ellis}, R.~S. 2013, \apjl, 762, L11

\bibitem[\protect\citeauthoryear{{Moore} et~al.}{{Moore}
  et~al.}{1999}]{moore1999}
{Moore}, B., {Ghigna}, S., {Governato}, F., {Lake}, G., {Quinn}, T., {Stadel},
  J.,  \& {Tozzi}, P. 1999, \apjl, 524, L19

\bibitem[\protect\citeauthoryear{{More} et~al.}{{More} et~al.}{2011}]{more2011}
{More}, S., {van den Bosch}, F.~C., {Cacciato}, M., {Skibba}, R., {Mo}, H.~J.,
  \& {Yang}, X. 2011, \mnras, 410, 210

\bibitem[\protect\citeauthoryear{{Moster} et~al.}{{Moster}
  et~al.}{2010}]{moster2010}
{Moster}, B.~P., {Somerville}, R.~S., {Maulbetsch}, C., {van den Bosch}, F.~C.,
  {Macci{\`o}}, A.~V., {Naab}, T.,  \& {Oser}, L. 2010, \apj, 710, 903

\bibitem[\protect\citeauthoryear{{Navarro}, {Frenk}, \& {White}}{{Navarro}
  et~al.}{1997}]{navarr1997}
{Navarro}, J.~F., {Frenk}, C.~S.,  \& {White}, S.~D.~M. 1997, \apj, 490, 493

\bibitem[\protect\citeauthoryear{{Newman} et~al.}{{Newman}
  et~al.}{2012}]{newman2012}
{Newman}, J.~A., et~al. 2012, ArXiv e-prints, arxiv:1203.3192

\bibitem[\protect\citeauthoryear{{Oh} et~al.}{{Oh} et~al.}{2011}]{oh2011}
{Oh}, S.-H., {Brook}, C., {Governato}, F., {Brinks}, E., {Mayer}, L., {de
  Blok}, W.~J.~G., {Brooks}, A.,  \& {Walter}, F. 2011, \aj, 142, 24

\bibitem[\protect\citeauthoryear{{Peng}}{{Peng}}{2010}]{peng2010}
{Peng}, C. 2010, in Bulletin of the American Astronomical Society, Vol.~42,
  Bulletin of the American Astronomical Society, 578

\bibitem[\protect\citeauthoryear{{Peter} et~al.}{{Peter}
  et~al.}{2013}]{peter2013}
{Peter}, A.~H.~G., {Rocha}, M., {Bullock}, J.~S.,  \& {Kaplinghat}, M. 2013,
  \mnras, 430, 105

\bibitem[\protect\citeauthoryear{{Pizagno} et~al.}{{Pizagno}
  et~al.}{2007}]{pizagn2007}
{Pizagno}, J., et~al. 2007, \aj, 134, 945

\bibitem[\protect\citeauthoryear{{Pontzen} \& {Governato}}{{Pontzen} \&
  {Governato}}{2012}]{pontze2012}
{Pontzen}, A.,  \& {Governato}, F. 2012, \mnras, 421, 3464

\bibitem[\protect\citeauthoryear{{Reyes} et~al.}{{Reyes}
  et~al.}{2012}]{reyes2012}
{Reyes}, R., {Mandelbaum}, R., {Gunn}, J.~E., {Nakajima}, R., {Seljak}, U.,  \&
  {Hirata}, C.~M. 2012, \mnras, 425, 2610

\bibitem[\protect\citeauthoryear{{Reyes} et~al.}{{Reyes}
  et~al.}{2011}]{reyes2011}
{Reyes}, R., {Mandelbaum}, R., {Gunn}, J.~E., {Pizagno}, J.,  \& {Lackner},
  C.~N. 2011, \mnras, 417, 2347

\bibitem[\protect\citeauthoryear{{Rocha} et~al.}{{Rocha}
  et~al.}{2013}]{rocha2013}
{Rocha}, M., {Peter}, A.~H.~G., {Bullock}, J.~S., {Kaplinghat}, M.,
  {Garrison-Kimmel}, S., {O{\~n}orbe}, J.,  \& {Moustakas}, L.~A. 2013, \mnras,
  430, 81

\bibitem[\protect\citeauthoryear{{Spergel} \& {Steinhardt}}{{Spergel} \&
  {Steinhardt}}{2000}]{sperge2000}
{Spergel}, D.~N.,  \& {Steinhardt}, P.~J. 2000, Physical Review Letters, 84,
  3760

\bibitem[\protect\citeauthoryear{{Springel} et~al.}{{Springel}
  et~al.}{2008}]{spring2008}
{Springel}, V., et~al. 2008, \mnras, 391, 1685

\bibitem[\protect\citeauthoryear{{Stark}, {McGaugh}, \& {Swaters}}{{Stark}
  et~al.}{2009}]{stark2009}
{Stark}, D.~V., {McGaugh}, S.~S.,  \& {Swaters}, R.~A. 2009, \aj, 138, 392

\bibitem[\protect\citeauthoryear{{Tacconi} et~al.}{{Tacconi}
  et~al.}{2010}]{taccon2010}
{Tacconi}, L.~J., et~al. 2010, \nat, 463, 781

\bibitem[\protect\citeauthoryear{{Teyssier} et~al.}{{Teyssier}
  et~al.}{2013}]{teyssi2013}
{Teyssier}, R., {Pontzen}, A., {Dubois}, Y.,  \& {Read}, J.~I. 2013, \mnras,
  429, 3068

\bibitem[\protect\citeauthoryear{{Tollerud} et~al.}{{Tollerud}
  et~al.}{2008}]{toller2008}
{Tollerud}, E.~J., {Bullock}, J.~S., {Strigari}, L.~E.,  \& {Willman}, B. 2008,
  \apj, 688, 277

\bibitem[\protect\citeauthoryear{{Weinberg} et~al.}{{Weinberg}
  et~al.}{2013}]{weinbe2013}
{Weinberg}, D.~H., {Bullock}, J.~S., {Governato}, F., {Kuzio de Naray}, R.,  \&
  {Peter}, A.~H.~G. 2013, ArXiv e-prints

\bibitem[\protect\citeauthoryear{{Wuyts} et~al.}{{Wuyts}
  et~al.}{2011}]{wuyts2011}
{Wuyts}, S., et~al. 2011, \apj, 738, 106

\bibitem[\protect\citeauthoryear{{Zolotov} et~al.}{{Zolotov}
  et~al.}{2012}]{zoloto2012}
{Zolotov}, A., et~al. 2012, \apj, 761, 71

\end{thebibliography}

\end{document}